# Ludwig-Soret microscopy with vibrational photothermal effect

Keiichiro Toda[1] and Takuro Ideguchi[1,*]

[1] Institute for Photon Science and Technology, The University of Tokyo, Tokyo, Japan

[*] Corresponding author: ideguchi@ipst.s.u-tokyo.ac.jp

**Abstract**

Vibrational microscopy provides label-free, bond-selective chemical contrast by detecting molecular vibrations, making it invaluable for biomedical research. While conventional methods rely on the direct detection of Raman scattering or infrared absorption, recently developed vibrational photothermal (ViP) microscopy achieves chemical contrast indirectly through refractive index (RI) changes. This indirect approach enables unique imaging capabilities beyond traditional chemical imaging. Here, we introduce a novel application of ViP microscopy: label-free intracellular thermophoretic (Soret) imaging, which visualizes biomolecular transport driven by temperature gradients. ViP-induced Soret (ViPS) imaging leverages a steady-state temperature distribution generated by optical heating through vibrational photothermal effect, combined with time-resolved RI imaging via optical diffraction tomography (ODT). Using ViPS imaging, we measured thermophoretic behavior in living COS7 cells, determining intracellular diffusion and Soret coefficients. Notably, we observed a reversed direction of molecular transport (negative Soret effect) in the cytoplasm compared to the nucleus, possibly driven by thermophoresis-induced diffusiophoresis. Furthermore, time-lapse imaging under $CO_2$-depleted conditions revealed a remarkable reduction in thermophoretic activity, suggesting glass formation during the dying process, likely due to polymer aggregation. ViPS imaging represents a new frontier in intracellular thermophoretic studies, expanding the capabilities of vibrational microscopy.

**Introduction**

Vibrational microscopy is widely recognized as a powerful tool for biomedical research due to its label-free and bond-selective capabilities[1,2]. Conventional vibrational microscopy achieves chemical contrast by directly reading molecular vibrational information through Raman scattering or infrared absorption. In contrast, newly emerging vibrational microscopy techniques provide chemical contrast indirectly through intermediate phenomena. For example, vibrational photothermal (ViP) microscopy, such as mid-infrared photothermal (MIP) microscopy, detects refractive index (RI) changes induced by the bond-selective ViP effect[3,4,5,6,7,8]. In this approach, RI changes are not directly caused by molecular vibrations but rather result from thermal expansion due to heat dissipation from molecular vibrations. Recent advancements in ViP microscopy have led to intensive development, demonstrating chemical imaging capabilities in cells comparable to state-of-the-art techniques such as stimulated Raman scattering (SRS) microscopy[9].

Vibrational microscopes that utilize intermediate phenomena, including ViP microscopes, have the potential to provide information beyond chemical contrast[10,11,12]. For example, we have recently demonstrated that time-resolved ViP imaging enables the measurement of intracellular thermal properties by detecting ViP-induced heat diffusion[12]. Since heat diffusion occurs on the microsecond timescale, time-resolved pump-probe ViP imaging made this experiment possible. RI changes can be induced not only by thermal expansion but also by other factors. As RI depends on molecular density (or cellular dry mass), RI image contrast varies with changes in molecular composition. We found that the ViP effect can drive directional molecular transport along a temperature gradient over a longer timescale. This phenomenon, known as thermophoresis or the Ludwig-Soret (or simply Soret) effect, has been widely studied in biophysics, primarily in vitro[13]. Previous studies have explored biomolecular mobility, providing insights into molecular crowding and interactions via hydration and charge[14,15]. These in vitro studies have also highlighted the potential for opto-thermal manipulation or sorting of intracellular biomolecules at significantly lower laser power than optical tweezers[16,17].

In this work, we propose and demonstrate intracellular label-free thermophoretic imaging of biomolecules using ViP microscopy, introducing a new capability of vibrational microscopy. Our ViP-induced Soret (ViPS) imaging is enabled by two key factors: (1) the creation of a steady-state, Gaussian-like temperature distribution induced by seconds-long optical heating via bond-selective continuous absorption of water and (2) time-resolved imaging of spatiotemporal RI variations caused by molecular movements (changes in dry mass concentration) using optical diffraction tomography (ODT). The optically induced temperature distribution within a cell drives seconds-long, slow molecular transport (thermophoresis), modifying the RI. This effect can be distinguished from the faster RI variations caused by thermal diffusion, which occur on micro- to millisecond timescales. Using ViPS imaging, we measured the thermophoretic behavior in the nucleus of a living COS7 cell, determining the diffusion and Soret coefficients to be approximately 4.6 $\mu m^2\ s^{-1}$ and 0.01 $K^{-1}$, respectively. This indicates that a 1K inhomogeneous intracellular temperature distribution induces a 1% change in dry mass for molecules with a molecular weight of approximately 100 kDa. Additionally, we observed molecular transport in the cytoplasm accruing in the opposite direction to that in the nucleus, likely due to the combined effects of thermophoresis and diffusiophoresis. Furthermore, time-lapse imaging under $CO_2$-depleted conditions revealed a significant reduction in thermophoretic movement, suggesting that glass formation occurs in dying cells due to polymer aggregation.

**Results**
**Working principle of ViPS imaging.**
ViPS imaging is built upon a ViP microscope, where a visible microscope detects RI changes induced by the bond-selective photothermal effect using mid- or near-infrared light and its subsequently occurring phenomena. Figure 1 categorizes ViP-induced phenomena based on their respective time scales. The first observable phenomenon is an RI change caused by local thermal expansion (photothermal effect), occurring on a nanosecond time scale (Fig. 1a). This mechanism is the basis of conventional ViP imaging, where chemical contrast is achieved through bond-selective ViP excitation[3]. The second phenomenon, thermal diffusion, occurs on a microsecond time scale (Fig. 1b). The dynamics of thermal diffusion can be visualized using time-resolved pump-probe imaging[12]. The third phenomenon, and the focus of this work, involves thermophoresis and diffusiophoresis, occurring in milliseconds to seconds (Fig. 1c). Thermophoresis refers to molecular transport driven by a temperature gradient, which occurs under continuous ViP excitation. When a spatially focused continuous ViP excitation beam is applied to a sample, it quickly establishes a steady-state temperature distribution, inducing molecular movement along temperature gradients over seconds. This molecular transport redistributes molecular composition, leading to changes in the RI map. A detailed theoretical description of thermophoresis is provided in the Methods section, "Thermophoresis (Soret effect) and diffusiophoresis".

The principle of ViPS imaging is described as follows. An induced RI change ($\Delta n$) under continuous heating can be expressed as

$$\Delta n = \Delta n_\rho + \Delta n_\sigma, \quad (Eq.\ 1)$$

where $\Delta n_\rho$ and $\Delta n_\sigma$ represent the RI changes caused by density ($\rho$) changes due to thermal expansion and dry mass concentration ($\sigma$) changes, respectively. ViPS imaging specifically measures $\Delta n_\sigma$. It is important to note that these two terms evolve on different time scales in cells. Our previous study revealed that the first term $\Delta n_\rho$ reaches a steady state within 5 ms after heating initiation due to rapid thermal diffusion[12]. Because $\Delta n_\rho$ can be considered constant at longer timescales, the temporal variation of the second term $\Delta n_\sigma$ can be separately extracted. In this work, we present 2D intracellular thermophoretic images based on depth-integrated dry mass concentration changes ($\Delta \sigma$ [fg/μm$^2$]) expressed as

$$\Delta \sigma = \frac{\int \Delta n_\sigma dz}{\alpha}, \quad (Eq.\ 2)$$

where $\alpha = 0.2 \times 10^{-3}$ [μm$^3$/fg] is the reported refractive index increment relative to dry mass concentration in cells[18].

The ability to measure $\Delta n_\rho$ at the onset of heating is crucial for precisely determining the induced temperature rise distribution, which is essential for quantitative analyses of diffusion and Soret coefficients. Under steady-state conditions, $\Delta n_\rho$ can be expressed as

$$\Delta n_\rho = \frac{dn}{dT} \Delta T_s, \quad (Eq.\ 3)$$

where $dn/dT$ and $\Delta T_s$ denote the thermo-optic coefficient and the steady-state temperature rise, respectively. Since the thermo-optic coefficient remains nearly constant within cells, the temperature rise distribution $\Delta T_s$ can be determined by measuring a $\Delta n_\rho$ image. By leveraging the time-resolved capability to distinguish thermal diffusion from molecular transport, ViPS microscopy enables the visualization of intracellular molecular thermophoresis from a single dataset of spatiotemporal RI variations.

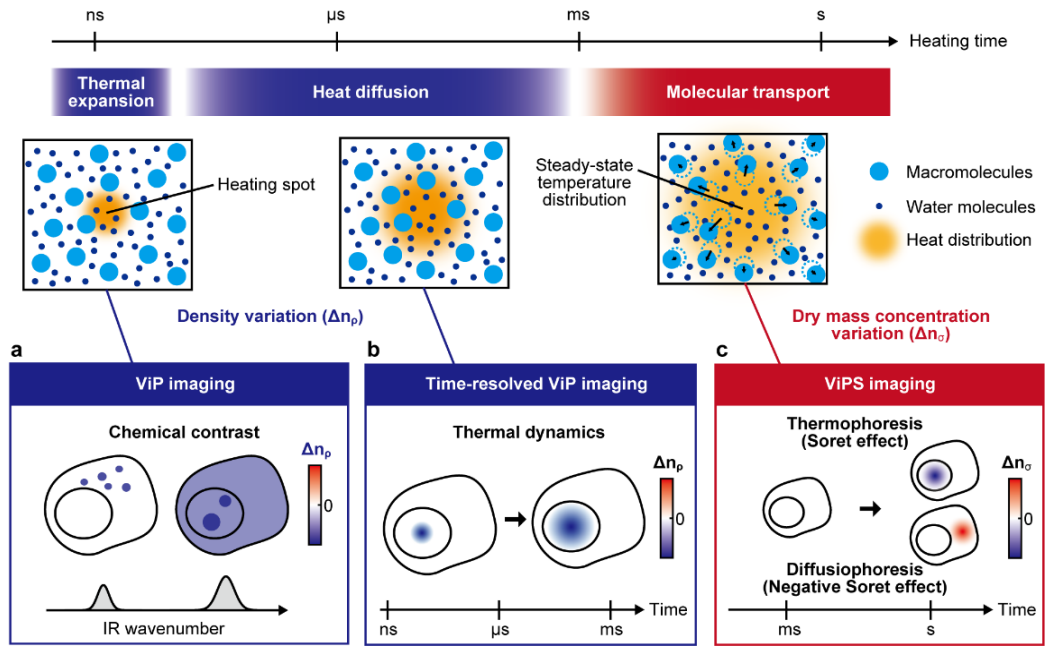

**Fig. 1 Principle of ViPS imaging illustrated with ViP-induced phenomena. a** ViP-induced thermal expansion on a nanosecond time scale, used in conventional ViP imaging to obtain chemical contrast. **b** ViP-induced heat diffusion on a microsecond timescale, visualized through time-resolved ViP imaging. **c** ViP-induced molecular transport occurring over milliseconds to seconds. Macromolecules migrate along temperature gradients, causing RI change. ViPS imaging captures thermophoresis (Soret effect) or thermophoresis-driven diffusiophoresis (negative Soret effect).

**ViPS microscopy system**

Our ViPS microscopy system is based on a ViP optical diffraction tomography (ViP-ODT) microscope[12]. Details of the system are provided in Methods "ViP-ODT system". In this study, we set the IR wavelength to an overtone band of water molecules, which are nearly uniformly distributed within cells. Continuous-wave (CW) IR light generates a Gaussian-like heat distribution, corresponding to the IR beam spot, with a full-width at half-maximum (FWHM) diameter of ~5 μm. The induced RI change is quantitatively measured using an ODT microscope with visible light. The ODT system, based on off-axis digital holography (DH)[19] with an azimuthally-scanned illumination scheme, provides 3-dimensional (3D) RI volumetric imaging with a spatial resolution of 250 nm laterally and 5 μm axially. The measurement temporal resolution is 20 ms, which is determined by the frame rate of the image sensor (50 Hz).

**Intracellular ViPS imaging and evaluation of Soret and diffusion coefficients.**

We conducted intracellular ViPS imaging of living COS7 cells. Firstly, a 2.4-second continuous heating was applied in the nucleus. Figure 2a illustrates the temporal variations of RI change at the center and the periphery of the heating spot, as indicated by the blue and red squares in Fig. 2b. Both curves exhibit instantaneous changes at heating onset, attributed to thermal expansion ($\Delta n_\rho$ in Eq. 1). Subsequently, gradual changes are observed, which can be explained by molecular transport ($\Delta n_\sigma$ in Eq. 1). The latter exhibits opposing behaviors, an increase at the center and a decrease at the periphery of the heating spot, indicating that molecular transport redistributes dry mass density within a boundary while mostly conserving total mass.

As discussed above, thermal expansion occurs rapidly, reaching a steady state typically within milliseconds, before molecular transport begins. Thus, the $\Delta n_\rho$ image can be extracted from the initial $\Delta n$ image captured immediately after the beginning of heating. Using Eq. 3, we derived the temperature change distribution ($\Delta T_s$), as shown in Fig. 2b, employing the thermo-optic coefficient of water (dn/dT = -1.0 × 10$^{-4}$ K$^{-1}$). The $\Delta n_\sigma$ images were obtained by subtracting the $\Delta n_\rho$ image from the $\Delta n$ images. Figure 2c presents an image of depth-integrated dry mass concentration change ($\Delta \sigma$), calculated using Eq. 2. The results clearly demonstrate that intracellular molecules migrate from higher to lower temperature regions, consistent with in vitro studies on biomolecular thermophoresis in aqueous environments[20]. A notable feature is that molecular transport primarily occurs inside the nucleus. The expelled molecules (red-colored in Fig. 2c) remain within the nucleus, likely because they are physically blocked by the nuclear membrane. We quantified the positive and negative dry mass changes within the nucleus and found that 85% of the expelled molecules at the heating spot remain inside the nucleus (see Supplementary Note 2 for details).

We evaluated the Soret ($S_T$) and diffusion ($D$) coefficients by comparing our experimental results with a numerical simulation of thermophoresis (see Methods "Numerical simulation of thermophoresis" for details on the calculation procedure). The temporal evolution of depth-integrated dry mass concentration change ($\Delta \sigma$) was simulated using the experimentally determined temperature rise distribution ($\Delta T_s$). Figure 2d shows the $\Delta \sigma$ maps at 0.1, 1, and 2.4 seconds after heating onset. Figure 2e presents its temporal evolution of $\Delta \sigma$ at the center of the heating region (averaged over the blue squared area shown in Fig. 2b). Through a systematic parameter search, we found that $S_T$ = 9.28 (± 0.10) × 10$^{-3}$ K$^{-1}$ and $D$ = 4.49 ± 0.10 μm²/s best reproduced the experimental results. The parameter search was performed in steps of 2 × 10$^{-5}$ K$^{-1}$ for $S_T$ and 0.02 μm²/s for $D$, minimizing the sum of squared differences between the experimental and simulated data points. Errors were evaluated using confidence intervals based on a residual increase of 10%.

The Soret coefficient quantifies molecular migration molecules induced by a 1 K temperature difference. The obtained value ($S_T$ = 9.28 × 10$^{-3}$ K$^{-1}$) indicates that ~1% of the molecules migrate after 2.4 seconds of heating with a 1 K temperature difference. Notably, our method can evaluate multiple $S_T$ values corresponding to molecules of different sizes from a single temporal RI evolution dataset with extended heating. This is because longer heating durations facilitate the transport of less mobile, larger molecules, which generally exhibit higher $S_T$ values.

The diffusion coefficient provides insight into the molecular size of thermophoretic biomolecules. Previous studies using fluorescence-based techniques, such as fluorescence recovery after photobleaching (FRAP), have reported intracellular diffusion coefficients for various biomolecules: 15-30 μm²/s for GFP (30 kDa)[21,22], 5.5 μm²/s for IgG

(160 kDa)[21], and 0.03-0.1 μm²/s for mRNA (~1000 kDa)[23]. The diffusion coefficient obtained in our experiment ($D$ = 4.6 μm²/s) is comparable to that of IgG, suggesting that the observed intracellular thermophoresis is primarily driven by biomolecules with molecular weights around 100 kDa or higher, such as proteins or small- to medium-sized RNAs. This inference aligns with prior knowledge of molecular confinement within the nucleus, where molecules larger than approximately 30 kDa cannot freely pass through the nuclear membrane[24].

To determine whether RNAs are dominant migrating molecules, we prepared RNA-reduced cells by treating them with Actinomycin D (See Methods "Preparation of biological samples" for details). Figure 2f presents the averaged intranuclear dry mass concentration at the center of the heating spot, comparing untreated and drug-treated cells. We observed a 13% reduction in dry mass following drug treatment. Given that ~10% of cellular dry mass is attributed to RNA in eukaryotic cells[22], this suggests that nearly all intracellular RNA was depleted, consistent with previous studies reporting a ~70-90% reduction in RNA levels following Actinomycin D treatment[25]. Figure 2g displays the thermophoresis-induced dry mass concentration change normalized by temperature increase ($-\Delta\sigma/\Delta T_s$), measured at the same region as Fig. 2f, with and without the RNA reduction. In RNA-reduced cells, thermophoresis signals decreased by 19%, indicating that approximately 20% of intracellular thermophoresis is driven by RNA migration. These results suggest that RNA is a major contributor to intracellular thermophoresis.

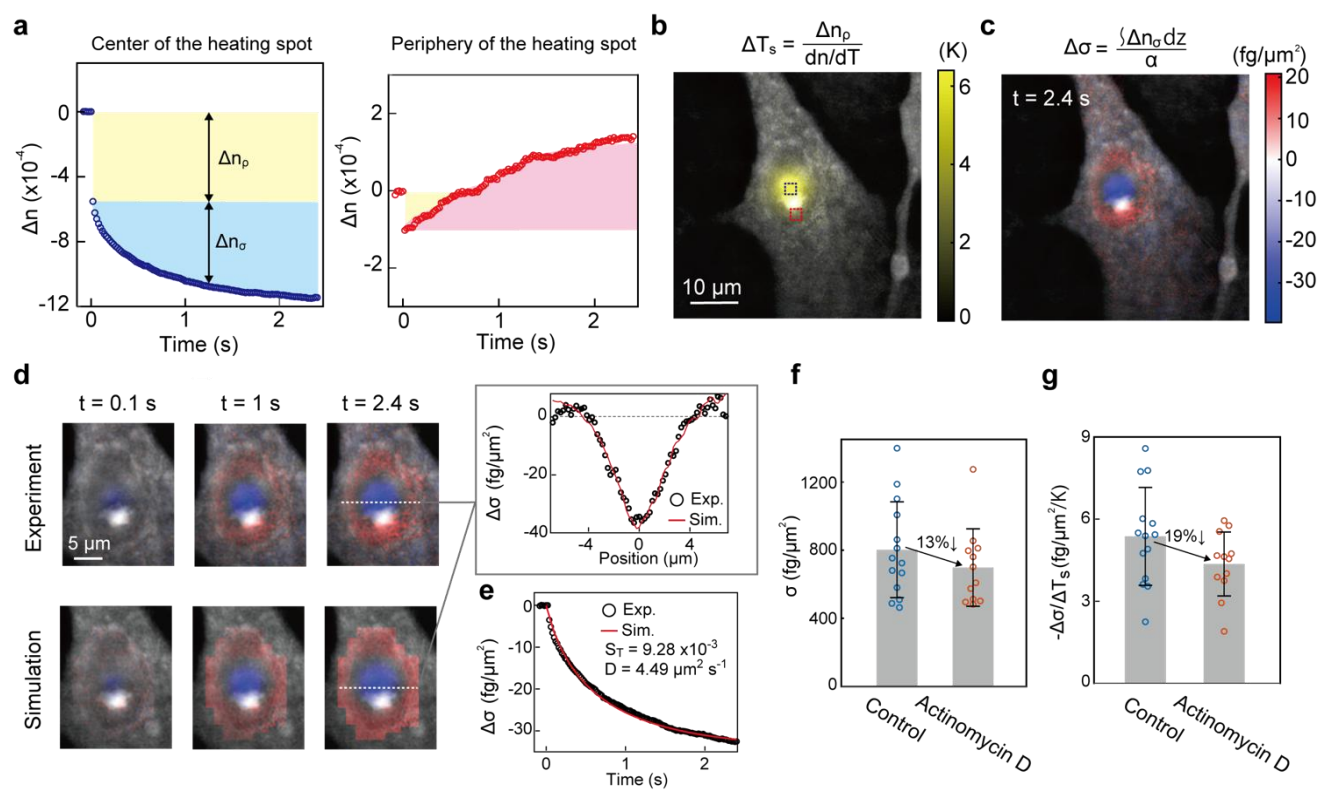

**Fig. 2 Intracellular thermophoretic imaging and evaluation of Soret and diffusion coefficients. a** Temporal variation in RI change ($\Delta n$) at the center and periphery of the heating spot (corresponding to the blue and red square regions in Fig. 2b). **b** Steady-state temperature rise map ($\Delta T_s$) derived from the $\Delta n_\rho$ map, measured 10 ms after heating onset. The temperature rise was determined by averaging a 4.2 μm-thick volumetric layer along the z-axis

within the cell. **c** Depth-integrated dry mass concentration change ($\Delta\sigma$) measured 2.4 seconds after heating onset. **d** Comparison of time-resolved $\Delta\sigma$ maps between experiment and simulation. The color scale matches that of Fig. 2c. The inset shows cross-sectional profiles along the white dotted lines. **e** Comparison of experimental and simulated results for the temporal evolution of $\Delta\sigma$ at the center of the heating spot (blue square region in Fig. 2b). **f** Depth-integrated dry mass concentration ($\sigma$) in cells with and without (control) Actinomycin D treatment, measured within a 2.2 μm × 2.2 μm area at the center of the heating spot. **g** Normalized depth-integrated dry mass concentration change ($-\Delta\sigma/\Delta T_s$) measured 2.4 seconds after heating onset. The number of measured cells was 14 for control and 12 for drug-treated.

**Site-specific intracellular thermophoresis in the nucleus and cytoplasm.**

We examined site-specific molecular transport by selectively heating different intracellular regions. In this study, we present the variation ratio of depth-integrated dry mass concentration ($\Delta\sigma/\sigma$), calculated by normalizing $\Delta\sigma$ with the RI image without heating ($\sigma$). Figure 3a presents the $\Delta T_s$ map, $\Delta\sigma/\sigma$ image at 2 seconds after heating onset, and the temporal evolution of $\Delta\sigma/\sigma$ at the center of the heating spot within the nucleus. For nuclear heating, we observed a reduction in dry mass at the center of the heating spot, while the surrounding region showed an increase in dry mass, consistent with the results shown in Fig. 2. Figure 3b shows the corresponding results for cytoplasmic heating, where molecular transport occurred in the opposite direction, moving toward the higher temperature region. To quantify this difference, we measured 12 cells and evaluated the magnitude and direction of molecular transport using $-\Delta\sigma/\sigma/\Delta T_s$, which corresponds to the Soret coefficient ($S_T$) under spherical symmetry. Figure 3c displays $-\Delta\sigma/\sigma/\Delta T_s$ at the center of the heating spot within the nucleus and cytoplasm, measured at 2.4 seconds, yielding values of $4.7 \times 10^{-3}$ and $-1.8 \times 10^{-2}$, respectively. These results clearly demonstrate a distinct difference in molecular transport direction between the nucleus and cytoplasm.

The opposite molecular transport observed in the cytoplasm, characterized by a negative Soret coefficient, can be explained by diffusiophoresis, which occurs in environments containing molecules of different sizes[26]. Under such conditions, smaller molecules undergo thermophoresis earlier than larger molecules. The resulting spatial concentration gradient of small molecules generates a force that pushes larger molecules opposite to the thermophoretic direction. When this force exceeds the thermophoretic effect on larger molecules, a negative Soret effect arises. To verify the rapid thermophoresis of small molecules, we analyzed $\Delta\sigma/\sigma$ images measured within the first 0.1 s of heating (Fig. 3d). At 0.04 s, molecules were expelled from the heating spot, exhibiting normal thermophoretic behavior. However, at 0.08 s, an opposite molecular transport toward the higher-temperature region began at the center of the heating spot. This observation suggests that early thermophoresis of small molecules triggers diffusiophoresis. Given that the cytoplasm contains a higher concentration of small molecules, such as metabolites, than the nucleus, it is reasonable that this effect was observed only in the cytoplasm.

An alternative explanation for the opposite molecular transport could be the optical tweezer effect; however, this was not the case in our experiment. To confirm this, we conducted the same experiment using a different IR wavelength

(1064 nm), which is off-resonance with water absorption. Under this condition, thermophoretic effects were suppressed, leaving only the optical tweezer effect. This experiment confirmed that the optical tweezer effect alone did not induce molecular transport. (see Supplementary Note 3 for details).

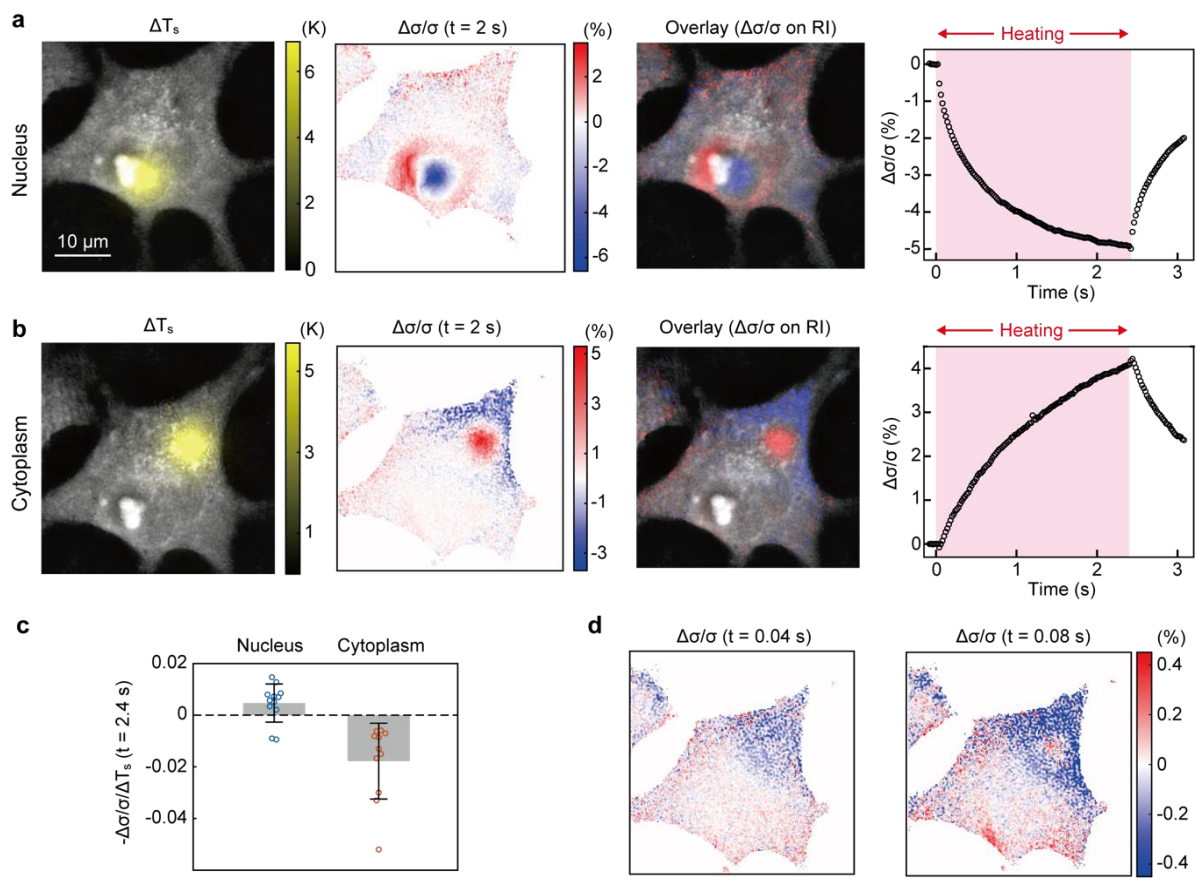

**Fig. 3 Site-specific thermophoretic behavior in the nucleus and cytoplasm. a** (left) Steady-state temperature rise map ($\Delta T_s$), (middle left) variation ratio of depth-integrated dry mass concentration ($\Delta\sigma/\sigma$), (middle right) overlay image ($\Delta\sigma/\sigma$ on RI), and (right) temporal evolution of $\Delta\sigma/\sigma$ at the center of the heating spot for nuclear heating. To visualize temporal evolution, $\Delta\sigma/\sigma$ values were averaged within a 2.2 μm × 2.2 μm area. **b** Corresponding images and the temporal evolution curve for cytoplasmic heating. **c** Evaluated values of $-\Delta\sigma/\sigma/\Delta T_s$, corresponding to the Soret coefficient under spherical symmetry, measured at 2.4 s for 12 cells (nucleus) and 11 cells (cytoplasm). **d** $\Delta\sigma/\sigma$ images at 0.04 and 0.08 s for cytoplasmic heating.

**Time-lapse observation of the thermophoretic reaction of dying cells**

The mobility of intracellular molecules is expected to vary depending on the cell's vital state, as cellular viscosity likely changes in dead cells due to molecular aggregation and other factors. To investigate these temporal changes, we conducted time-lapse imaging of cellular morphology and thermophoretic behavior under room temperature conditions without $CO_2$ supply. Figure 4a presents RI and $\Delta\sigma/\sigma$ images measured 2.4 seconds after heating onset within the nucleus, comparing representative images captured at the start of time-lapse measurement (0 h) and 5 hours later. The RI image at 5 hours reveals vivid contrasts, indicating localized regions of high dry mass concentration along intracellular structures, whereas the 0-hour image exhibits smoother contrasts. Similarly, the $\Delta\sigma/\sigma$ images show distinct differences: at 0 hours, significant molecular mobility is observed, whereas at 5 hours, thermophoresis is absent. Figure 4b illustrates the temporal evolution of $\Delta\sigma/\sigma$ at the center of the heating spot, clearly showing the absence of molecular transport in the 5-hour image. Figure 4c presents the time-lapse variations of $\Delta\sigma/\sigma$ at 2.4 s after the heating onset within the nucleus for four different cells, all exhibiting a progressive decline in mobility over several hours. This observation, combined with the localized dry mass distribution visible in the RI image, suggests that glass-like formation may occur in dying cells, possibly due to polymer aggregation[27].

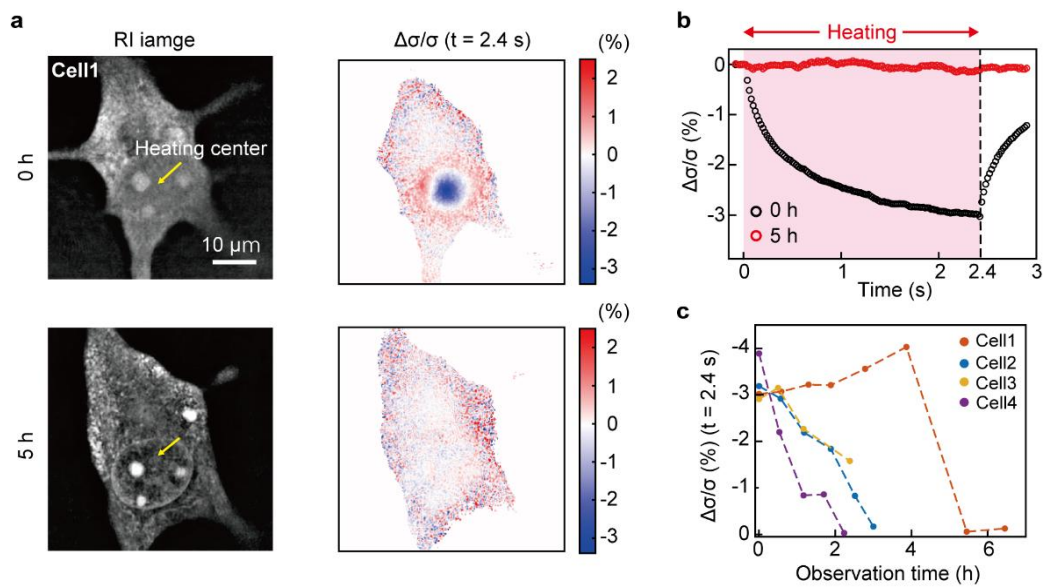

**Fig. 4 Time-lapse observation of intracellular thermophoretic behavior. a** RI and $\Delta\sigma/\sigma$ images captured at 0 hours and 5 hours. The $\Delta\sigma/\sigma$ images were acquired 2.4 s after heating onset. **b** Temporal evolution of $\Delta\sigma/\sigma$ at the center of the heating spot, averaged over a 2.2 μm × 2.2 μm region. **c** Time-lapse variations of $\Delta\sigma/\sigma$, measured 2.4 s after heating onset within the nucleus for four different cells.

## Discussion

To the best of our knowledge, intracellular thermophoresis has been previously explored in only one study, which utilized a fluorescence imaging technique[28]. This study examined the reduction in the diffusion coefficient within the cellular environment compared to an aqueous environment by introducing fluorescent dyes (BCECF) and 21-mer DNA molecules into living cells. However, fluorescence imaging is limited to visualizing specific molecular species, making it challenging to obtain a comprehensive view of molecular density dynamics. Additionally, fluorescence imaging cannot directly measure temperature gradient maps, restricting the precise analysis of diffusion and Soret coefficients. In contrast, ViPS microscopy overcomes these limitations by simultaneously capturing intracellular molecular density changes and temperature rise distributions through RI changes, all within a single, label-free measurement. These advanced capabilities provide richer information and enable quantitative analyses of intracellular thermophoretic dynamics.

There is room for improvement in our current ViPS imaging system. One key area is enhancing the detectable minimum $\Delta\sigma/\sigma$. Currently, this limitation is not due to RI noise in the ODT system (~$10^{-5}$ without averaging, corresponding to $\Delta\sigma/\sigma$ ~0.1%) but rather the random movement of small particles within the intracellular environment, particularly in the cytoplasm. This issue arises from taking RI differences between frames with a time interval exceeding one second. In our experiments, the noise values, evaluated as the standard deviation of $\Delta\sigma/\sigma$ images at 2 seconds without heating, were approximately 0.15% in the nucleus and 0.6% in the cytoplasm. A potential solution to reduce random molecular motion noise is to perform a correlation analysis of $\Delta n$ images with and without IR heating. This approach takes advantage of the directional nature of thermophoretic transport, distinguishing it from random molecular movement and thereby improving measurement precision. Another promising direction is bond-specific thermophoretic imaging, which leverages the chemical contrast capability of ViP imaging. By incorporating bond-selective chemical imaging using MIR light, it would be possible to selectively observe the thermophoresis of specific molecular species, such as proteins, lipids, or DNA. This advancement would enable the spatiotemporal measurement of concentration changes for individual molecules, offering deeper insights into their thermophoretic behaviors.

Finally, we outline several future prospects. Thermophoresis can be utilized to regulate cellular functions by manipulating biomolecules. Our study indicates that the molecules contributing to the observed thermophoresis are likely proteins and small- to medium-sized RNAs. Thermophoresis-induced molecular concentration variations may potentially trigger liquid-liquid phase separation (LLPS), which could be observable within cellular environments using our microscope. Another intriguing possibility is the manipulation of mRNA or DNA via thermophoresis, potentially leading to modifications in cellular states. Monitoring this phenomenon through ViPS imaging could pave the way for precise control of cellular dynamics. However, the diffusion coefficients of these large molecules are two orders of magnitude lower than those examined in this study. Consequently, extended heating durations on the order of approximately 100 seconds may be necessary to detect such effects.

The observed reverse molecular transport in the cytoplasm may arise from more complex phenomena beyond simple thermophoresis-induced diffusiophoresis. In our experiment, approximately 0.5% of the total dry mass concentration migrated toward the lower temperature side during the initial heating stage within 0.08 s. This corresponds to a solute concentration of ~0.1% in water, calculated based on the cellular solute concentration of ~0.3 g/mL[22]. This concentration is an order of magnitude lower than that observed in a previous in vitro study[26], where reverse molecular transport of DNA was induced by thermophoresis of PEG solutes with a solute concentration change of ~1%. However, the cellular environment differs significantly from in vitro dilute solutions due to factors such as the excluded volume effect, hydrophobic/hydrophilic interactions, and variations in chemical potential. Furthermore, other intracellular phenomena, such as LLPS or temperature-sensitive molecular reactions, may also contribute to thermophoresis-driven reverse molecular transport under relatively small solute concentration changes. To further investigate this phenomenon, molecular-specific thermophoretic imaging with ViP chemical contrast or simultaneous fluorescence imaging can be employed. These techniques can detect variations segregated by molecular size and species, offering greater specificity in identifying the underlying mechanisms and advancing our understanding of the complex cellular environment.

Our time-lapse thermophoretic imaging of dying cells has provided profound insights into the mystery of how living cells maintain fluidity without aggregation, despite being densely packed with proteins. It has been hypothesized that ATP plays a critical role in sustaining intracellular fluidity in living cells[29]. The random motion of intracellular particles, observed via 3D RI imaging with ODT, is believed to reflect this phenomenon[30]. Investigating changes in thermophoretic behavior using our ViPS imaging could offer a clearer understanding of the mechanisms underlying intracellular fluidity. This approach may uncover the specific role of ATP and its impact on maintaining the dynamic nature of the intracellular environment.

**Reference**

1. Wrobel, T. P *et al*. Infrared Spectroscopic Imaging Advances as an Analytical Technology for Biomedical Sciences. *Anal. Chem.* **90**, 1444–1463 (2018).
2. Shipp, D. W. *et al*. Raman spectroscopy: techniques and applications in the life sciences. *Adv. Opt. Photonics* **9**, 315 (2017).
3. Bai, Y. *et al*. Bond-selective imaging by optically sensing the mid-infrared photothermal effect. *Sci. Adv.* **7**: eabg1559 (2021).
4. Zhang, D. *et al*. Depth-resolved mid-infrared photothermal imaging of living cells and organisms at submicrometer spatial resolution. *Sci. Adv.* **2**: e1600521 (2016).
5. Fu, P. *et al*. Super-resolution imaging of non-fluorescent molecules by photothermal relaxation localization microscopy. *Nat. Photonics* **17**, 330–337 (2023).
6. Yin, J. *et al.* Video-rate mid-infrared photothermal imaging by single-pulse photothermal detection per pixel. *Sci. Adv.* **9**, eadg8814 (2023).
7. Ishigane, G. *et al.* Label-free mid-infrared photothermal live-cell imaging beyond video rate. *Light Sci. Appl.* **12**, 174 (2023).
8. Tamamitsu, M. *et al.* Mid-infrared wide-field nanoscopy. *Nat. Photonics* **18**, 738-743 (2024).
9. Freudiger, C. W. *et al.* Label-Free Biomedical Imaging with High Sensitivity by Stimulated Raman Scattering Microscopy. *Science* **322**, 1857–1861 (2008).
10. Samolis, P. D. *et al.* Label-free imaging of fibroblast membrane interfaces and protein signatures with vibrational infrared photothermal and phase signals. *Biomed. Opt. Express* **12**, 5400 (2021).
11. Yin, J. *et al.* Nanosecond-resolution photothermal dynamic imaging via MHz digitization and match filtering. *Nat. Commun.* **12**, 1–11 (2021).
12. Toda, K. *et al.* Label-free mid-infrared photothermal microscopy revisits intracellular thermal dynamics : what do fluorescent nanothermometers measure ? *arXiv* 2406.16265 (2024).
13. Duhr, S. *et al.* Why molecules move along a temperature gradient. *PNAS* **103**, 19678–19682 (2006).

14. Baaske, P. *et al.* Optical thermophoresis for quantifying the buffer dependence of aptamer binding. *Angew. Chem., Int. Ed.* **49**, 2238–2241 (2010).
15. Wienken, C. J. *et al.* Protein-binding assays in biological liquids using microscale thermophoresis. *Nat. Commun.* **1**, 100 (2010).
16. Chen, Z. *et al.* Heat-Mediated Optical Manipulation. *Chem. Rev.* **122**, 3122–3179 (2022).
17. Kollipara, P. S. *et al.* Hypothermal opto-thermophoretic tweezers. *Nat. Commun.* **14**, 1–9 (2023).
18. Popescu, G. *et al.* Optical imaging of cell mass and growth dynamics. *Am. J. Physiol. - Cell Physiol.* **295**, 538–544 (2008).
19. Park, Y. K. *et al.* Quantitative phase imaging in biomedicine. *Nat. Photonics* **12**, 578–589 (2018).
20. Fukuyama, T. *et al.* Opto-thermal diffusiophoresis of soft biological matter: from physical principle to molecular manipulation. *Biophys. Rev.* **12**, 309–315 (2020).
21. Arrio-Dupont, M. *et al.* Translational diffusion of globular proteins in the cytoplasm of cultured muscle cells. *Biophys. J.* **78**, 901–907 (2000).
22. Milo, R. *Cell Biology by the Numbers. 1st edn. (New York: Garland Science, 2015).*
23. Shav-Tal, Y. *et al.* Dynamics of single mRNPs in nuclei of living cells. *Science* **304**, 1797–1800 (2004).
24. Dirk, G. *et al.* Transport between the cell nucleus and the cytoplasm. *Annu. Rev. Cell Dev. Biol.* **15**, 607–660 (1999).
25. Perry, R. P. *et al.* Inhibition of RNA synthesis by actinomycin D: Characteristic dose-response of different RNA species. *J. Cell. Physiol.* **76**, 127–139 (1970).
26. Maeda, Y. *et al.* Thermal separation: Interplay between the soret effect and entropic force gradient. *Phys. Rev. Lett.* **107**, 1–4 (2011).
27. Nishizawa, K. *et al.* Universal glass-forming behavior of in vitro and living cytoplasm. *Sci. Rep.* **7**, 1–12 (2017).
28. Reichl, M. *et al.* Thermophoretic manipulation of molecules inside living cells. *J. Am. Chem. Soc.* **136**, 15955–15960 (2014).
29. Umeda, K. *et al.* Activity-dependent glassy cell mechanics II: Nonthermal fluctuations under metabolic activity. *Biophys. J.* **122**, 4395-4413 (2023).
30. Hugonnet, H. *et al.* Improving Specificity and Axial Spatial Resolution of Refractive Index Imaging by Exploiting Uncorrelated Subcellular Dynamics. *ACS Photonics* **11**, 257–266 (2024).
31. Maeda, Y. *et al.* Effects of long DNA folding and small RNA stem-loop in thermophoresis. *PNAS* **109**, 17972–17977 (2012).

## Methods

**Thermophoresis (Soret effect) and diffusiophoresis.**

A spatial temperature gradient in a solution induces unidirectional molecular transport. The total flux $J$ of molecules under the temperature gradient $\nabla T$ can be described by molecular diffusion and thermophoretic transport, expressed as

$$J = -D\nabla c(\boldsymbol{x},t) - D_T \nabla T(\boldsymbol{x},t)c(\boldsymbol{x},t), \qquad (\text{Eq. 4})$$

where $D$, $D_T$, $c$, $\boldsymbol{x}$, and $t$ denote the diffusion coefficient, thermophoretic mobility, solute concentration, three-dimensional spatial coordinates, and time, respectively[20]. The temporal variation of the solute concentration distribution $c(\boldsymbol{x},t)$ is governed by

$$\frac{\partial c(\boldsymbol{x},t)}{\partial t} = -\nabla \cdot J = D\nabla \cdot \left[\left(\nabla + S_T \nabla T(\boldsymbol{x},t)\right)c(\boldsymbol{x},t)\right], \qquad (\text{Eq. 5})$$

where the Soret coefficient, defined as $S_T = D_T/D$, represents the magnitude and direction of $\nabla c$ arising from the temperature gradient.

The transport of biomolecules, such as protein, DNA, and RNA (hereafter referred to as solute 1), within an aqueous environment is known to be directed toward the lower-temperature side $(S_T > 0)$[20]. However, this direction reverses upon the introduction of another, smaller solute (hereafter referred to as solute 2)[26,31]. This phenomenon finds its origin in diffusiophoresis, which arises from the concentration gradient of solute 2 induced by thermophoresis. This gradient drives solute 1 in the opposite direction. In such case, the net flux of solute 1 can be expressed as

$$J_1 = -D\nabla c_1(\boldsymbol{x},t) - D_T \nabla T(\boldsymbol{x},t)c_1(\boldsymbol{x},t) + c_1(\boldsymbol{x},t)u(\boldsymbol{x},t), \qquad (\text{Eq. 6})$$

where $c_1$ and $u$ represent the concentration and diffusiophoretic velocity of solute 1, respectively. Note that $u$ depends on the concentration gradient of solute 2. This phenomenon is treated as analogous to thermophoresis but with a negatively signed Soret coefficient.

**ViP-ODT system.**
A detailed schematic of the ViP-ODT system is presented in Fig. S1 of Supplementary Note 1. A homemade 10-ns Ti:Sapphire laser with a 1 kHz repetition rate generates visible probe pulses at 705 nm, which are introduced into a Mach-Zehnder interferometer via a single-mode optical fiber. In the interferometer's sample arm, the probe pulses pass through beam-steering optics, consisting of a grating with a line density of 70 lines per millimeter (46-068, Edmund), an aperture, relay lenses, and a reflective objective lens (5007-000, Beck Optronic Solutions) positioned in a 4-f configuration. This setup illuminates the sample at an angled incidence with a numerical aperture (NA) of 0.58. The illumination angle can be adjusted in discrete steps by rotating the grating through 10 different angles, with the aperture synchronously rotating to transmit only the first-order diffraction beam. The FWHM illumination area on the sample is 110 μm, corresponding to a 1/11.8 times demagnification from the grating. For holographic measurement, the sample image is magnified by a factor of 167 onto the image sensor plane (Q-2HFW, Adimec) using an objective lens (LCPLFLN100XLCD, Olympus) and relay lenses. The reference light is directed to the image sensor in an off-axis configuration after matching the optical path length, beam size, and polarization to those of the object light.

We used a 1,456-nm CW laser diode (LD) (BL1456-PAG500, Thorlabs) for IR heating (CW-IR), which was electronically modulated using an LD driver (see Supplementary Note 1 for the timing chart of ViPS imaging). The rise time of the IR laser is sufficiently short compared to the measurement frame rate, as shown in Fig. S2. The CW-IR beam was spatially combined with the visible beam via a dichroic mirror and focused onto the sample through the reflective objective lens. The IR power used in ViPS imaging was ~2 mW.

**Preparation of biological samples.**

COS7 cells were cultured on a $CaF_2$ substrate (C20SQ-0.5, Pier Optics) using high-glucose Dulbecco's modified Eagle medium (DMEM) containing L-glutamine, phenol red, and HEPES (FUJIFILM Wako). The medium was supplemented with 10% fetal bovine serum (Cosmo Bio) and 1% penicillin-streptomycin-L-glutamine solution (FUJIFILM Wako). The cells were maintained at 37°C in a 5% $CO_2$ atmosphere. For imaging, the COS7 cells were sandwiched between the substrates with a 10-μm spacer (PCIMRS, MISUMI). The environmental temperature during measurements was 24°C.

To reduce intracellular RNA levels for the experiment shown in Fig. 2f and 2g, Actinomycin D (FUJIFILM Wako) was added to the cultured cells at a concentration of 2.0 μg/mL. Time-lapse imaging was then performed for six hours following the treatment.

**Numerical simulation of intracellular thermophoresis.**

To calculate the temporal evolution of the thermophoresis-induced change in depth-integrated dry mass concentration ($\Delta\sigma$), we performed a two-dimensional diffusion simulation under a temperature gradient. A numerical program, implemented in Python with GPU acceleration, was developed based on the Forward Time Centered Space (FTCS) method. The pixel pitch was set to 207 nm to match the ODT measurement. For this calculation, a slightly modified version of Eq. 5 was used, expressed as

$$\frac{\partial(\Delta\sigma(x,t))}{\partial t} = D\nabla^2\big(\Delta\sigma(x,t)\big) + DS_T\nabla \cdot [\nabla T_s(x)\{\sigma(x) + \Delta\sigma(x,t)\}]. \tag{Eq. 7}$$

Here, the diffusion term associated with the steady-state inhomogeneous dry mass distribution prior to heating (i.e., $\nabla^2(\sigma(x))$) is excluded, based on the assumption that the diffusion forces from this steady-state distribution are balanced by forces generated by factors other than the temperature gradient.

To ensure the conservation of the dry mass within the nuclear boundary, as evidenced by the experimental results shown in Fig. 2, the pixels at the edge of the nuclear region were constrained by the following equation

$$\nabla(\Delta\sigma(x,t)) + S_T\nabla T_s(x)\{\sigma(x) + \Delta\sigma(x,t)\} = 0. \tag{Eq. 8}$$

Additionally, to prevent molecules from diffusing out of the region, the gradients of dry mass concentration and temperature were set to zero both inside and outside the boundary. The nuclear region was estimated from the raw RI image of the cell (see Supplementary Note 2 for details).

**Data availability**

The data provided in the manuscript is available from T.I. upon reasonable request.

**Acknowledgments**

This work was financially supported by Japan Society for the Promotion of Science (23H00273), JST FOREST Program (Grant Number JPMJFR236C, Japan), Precise Measurement Technology Promotion Foundation. We thank Kohki Okabe for invaluable discussions regarding the interpretation of our results and guidance in sample preparation, and Zicong Xu for critically reading the manuscript.

**Author contributions**

K.T. developed the ViP-ODT system, performed the experiments, and analyzed the experimental data. T.I. supervised the work. K.T. and T.I. wrote the manuscript.

**Competing interests**

K.T. and T.I. are inventors of patents related to ViP-ODT.

# Supplementary Information for

# Ludwig-Soret microscopy with vibrational photothermal effect

Keiichiro Toda[1], and Takuro Ideguchi[1,*]

[1] Institute for Photon Science and Technology, The University of Tokyo, Tokyo, Japan

[*] Corresponding author: ideguchi@ipst.s.u-tokyo.ac.jp

**Supplementary Note 1: Detailed schematic of ViP-ODT system.**

Figure S1 represents the schematic of our ViP-ODT system. Details are described in Methods "ViP-ODT" in the main manuscript.

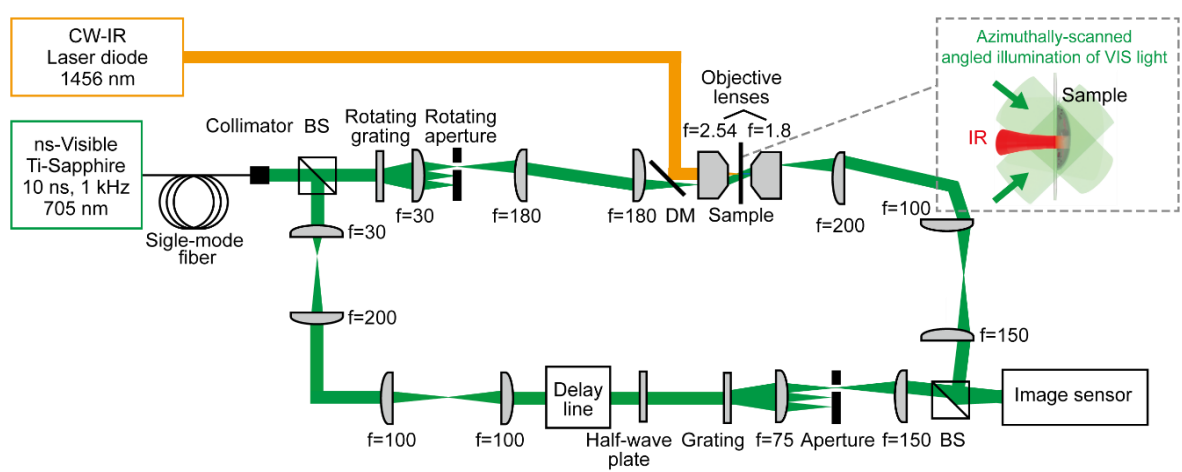

**Fig. S1 Detailed schematic of ViP-ODT.** DM: Dichroic mirror, BS: Beamsplitter.

Figure S2 shows the timing chart of the system. The timing of the IR heating, VIS pulse, and sensor exposure was electrically controlled using a function generator (DG1022Z, RIGOL). We repeatably applied a 2.4-s heating phase followed by 10.2-s cooling phase to capture the temporal variation of the phase images for each VIS illumination angle. The sensor exposure begins shortly (~100 ms) before heating. Using the measured data from multiple angles, the temporal variation of $\Delta n$ was reconstructed. The cooling duration (10.2 s) was determined based on the time scales of thermal diffusion and molecular transport, ensuring that the observed cell system returns to its initial state before the subsequent heating phase with a different VIS illumination angle begins.

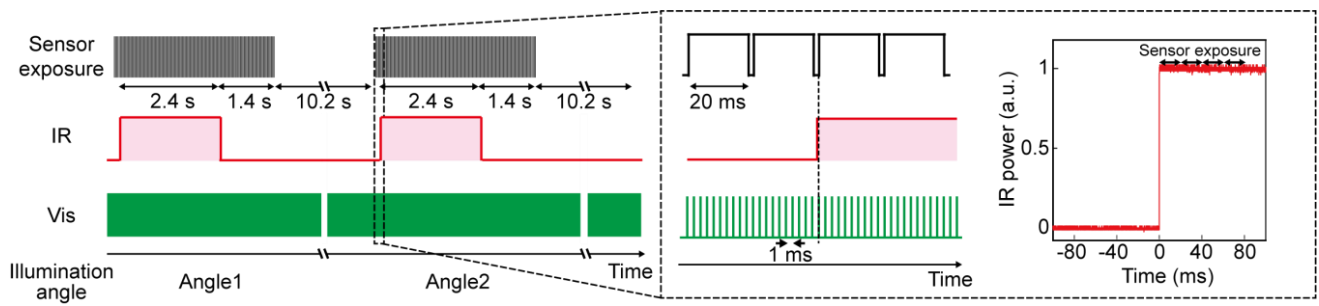

**Fig. S2 Timing chart of ViP-ODT.**

**Supplementary Note 2: Calculation of the confinement ratio of thermophoresis-induced expelled molecules within the nucleus.**

The confinement of thermophoresis-induced expelled molecules within the nucleus was confirmed through the following procedure. First, the nuclear region was identified based on the nuclear boundary observed in the cellular RI image (Fig. S3a). Next, the spatial integration of $\Delta\sigma$ across the nuclear region was performed to calculate the total dry mass change within the nucleus (Fig. S3b). Additionally, the total negative dry mass change, corresponding to molecules expelled from the heating spot, was quantified as shown in Fig. S3c. The ratio of these values indicates the fraction of molecules that escaped from the nucleus among those displaced by thermophoresis.

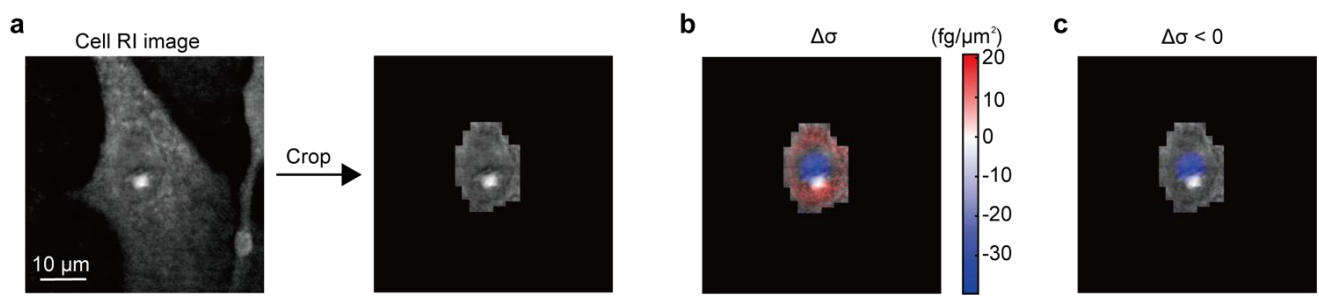

**Fig. S3 Calculation of $\Delta\sigma$ within the nuclear region. a** Nuclear region estimated from the cellular RI image. **b** $\Delta\sigma$ within the nuclear region. **c** Negative $\Delta\sigma$ within the nucleus.

**Supplementary Note 3: Validation of contribution of optical tweezer effect to the opposite transport observed within the cytoplasm.**

To validate the contribution of the optical tweezer effect independently of the thermophoresis effect, we conducted a comparative experiment using a 1064-nm laser, which is off-resonance with water absorption, thereby minimizing the contribution of thermophoresis to negligible levels. We compared the temporal variations of $\Delta\sigma/\sigma$ using two different pump wavelengths: 1450 nm (used in this study to induce thermophoresis via strong water absorption) and 1064 nm. The spot size was adjusted to be identical.

Figure S4a shows the cross-sectional spatial profiles of the temperature rise ($\Delta T_s$) caused by water absorption. To measure the weak signal at 1064 nm, we used a five times higher power (9 mW) (shown in the blue line). The magnitude of $\Delta T_s$ at 1064 nm is 70 times smaller than at 1450 nm under the assumption of identical illumination power with linear scalability, indicating a significant reduction in absorption.

Next, we set the illumination power to 1.3 and 1.8 mW for 1450 and 1064 nm, respectively, and measured the temporal variation of $\Delta\sigma/\sigma$. Figure S4b presents a comparison of the temporal variations of $\Delta\sigma/\sigma$ for both wavelengths, clearly demonstrating that $\Delta\sigma/\sigma$ induced by the 1064-nm pump light is substantially lower (negligible in this measurement) than that observed with the 1450-nm pump, despite both cases introducing comparable optical tweezer effects. This result suggests that the optical tweezer effect is not the primary contributor to $\Delta\sigma/\sigma$.

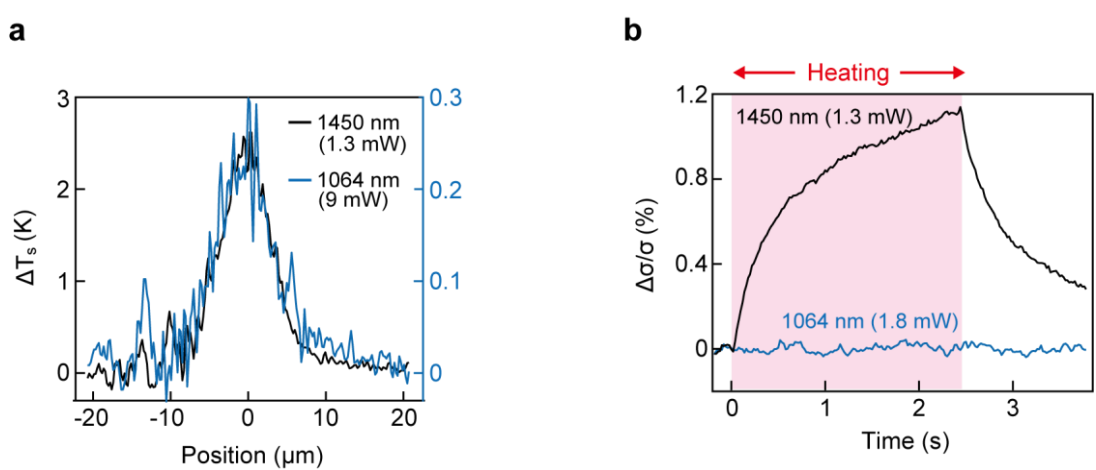

**Fig. S4 Comparison of $\Delta\sigma/\sigma$ between different pump wavelengths. a** $\Delta T_s$ cross-sectional spatial profiles and **b** temporal variation of $\Delta\sigma/\sigma$ for pump wavelengths of 1450 nm (black) and 1064 nm (blue). To generate the blue profile shown in **a**, a 1064 nm pump light with five times higher power (9 mW) was used to enhance the signal-to-noise ratio. For the measurement shown in **b,** the power levels were set to 1.8 mW for 1064 nm and 1.3 mW for 1450 nm.